# Neutron-gamma discrimination of CsI(Na) crystals for dark matter searches

Xilei Sun[a*], Junguang Lu[a], Tao Hu[a], Li Zhou[a], Jun Cao[a], Yifang Wang[a], Liang Zhan[a], Boxiang Yu[a], Xiao Cai[a], Jian Fang[a], Yuguang Xie[a], Zhenghua An[a], Zhigang Wang[a], Zhen Xue[a], Qiwen Lu[b], Feipeng Ning[a], Yongshuai Ge[a], Yingbiao Liu[a]

[a]*Institute of High Energy Physics, CAS, Beijing 100049, China*

[b]*Shanxi University, Taiyuan 030006, China*

**Abstract**

The luminescent properties of CsI(Na) crystals are studied in this report. By using a TDS3054C oscilloscope with a sampling frequency of 5 GS/s, we find out that nuclear recoil signals are dominated by very fast light pulse with a decay time of ~20 ns, while γ-ray signals have a decay time of ~600 ns. The wavelength of nuclear recoil and γ-ray signals are also different. The study of n/γ separation shows that the rejection factor can reach an order of $10^{-7}$ with signal efficiency more than 80% at an equivalent electron recoil energy of 20 keV or more. Such a property makes CsI(Na) an ideal candidate for dark matter searches.

## 1. Introduction

It is now commonly believed that dark matter accounts for 23% of the energy of the universe [1]. The best-motivated candidate of dark matter is the Weakly Interacting Massive Particles (WIMPs) with a mass between a few tens of GeV to a few TeV. In several well-motivated theories beyond the Standard Model, such as supersymmetry (SUSY) [2], the mass and interaction strength of WIMPs are predicted. WIMPs could scatter on terrestrial nuclei, leading to an exponential energy-transfer spectrum with a mean energy of about tens of keVr (nuclear recoil energy) [3]. The event rate is estimated to be below 0.1 events per kilogram of target per day, corresponding to a spin-independent WIMP-nucleon scattering cross section of $O$ ($10^{-43}$) cm$^2$ and a local dark matter halo density of 0.3 GeV/cm$^3$ in the neighborhood of our solar system [4]. Since the signal rate is so low, the main challenge for the direct detection of WIMPs is the identification of nuclear recoils from backgrounds such as electron recoils produced by γ-rays and betas from radioactive decays. There are generally two kinds of backgrounds: internal ones from impurities of detector materials and external ones from environment. While external backgrounds can be suppressed efficiently by well-designed shielding and active anti-coincidence detectors, internal radioactive background must be dealt with by a special selection of detector material. In fact, one can either select very pure detector materials, usually at a high cost, or detectors with the capability to distinguish γ-ray backgrounds from the nuclear recoil signal.

On the other hand, the total spin-independent scattering cross section of WIMPs on nuclei is roughly proportional to the square of the target atomic mass number (A$^2$) [5]. Among commonly used detector materials, CsI has the highest event rate in the recoil energy region <40 keVr as illustrated in Fig. 1 [6]. However, it is known that CsI(Tl) crystals, even specially produced with low backgrounds, consist of radioactive isotopes such as $^{137}$Cs and $^{87}$Rb, normally at a concentration of 2 mBq/kg and 1 ppb, respectively [7]. The total internal backgrounds (γ-rays and betas) are about $4\times10^6$ evt/100kg/year, limiting CsI crystal as a massive dark matter detector.

---

[*] Corresponding author. Tel.: +86-010-8823-5846; Fax: +86-010-8823-6423; E-mail: sunxl@ihep.ac.cn (X. Sun).

In this paper, we report our findings that CsI(Na) has very different responses to neutrons and γs. Neutrons deposit energy in the crystal via nuclear recoil in the same way as WIMPs. Therefore, internal γ backgrounds can be discriminated from the dark matter events. In fact, the scintillating waveforms induced by neutrons and γs in CsI(Na) crystals are very different, resulting a great Pulse Shape Discrimination (PSD) capability. The wavelength of scintillation induced by neutron and γ are also different, helping the discrimination by using a wavelength filter. A short discussion about the scintillating mechanism of such abnormal phenomena is given at the end of this paper. For convenience, the electron recoil equivalent energy is in units of keVee and the nuclear recoil energy is in units of keVr.

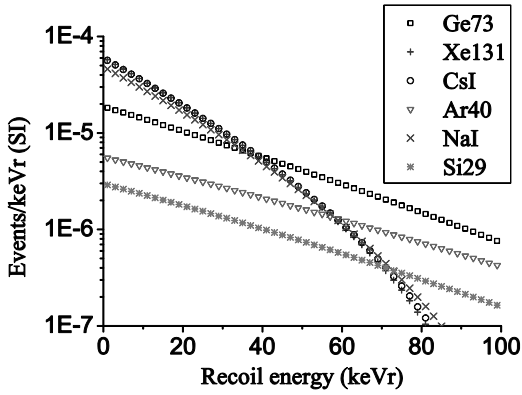

Fig. 1. Event rate of WIMP-nuclei scattering as a function of recoil energy for different target materials. The assumed WIMP mass is 50 GeV. WIMP-nucleon cross section is $10^{-7}$ pb and the exposure is 10 kg days.

**2. Experimental set-up**

The CsI(Na) crystal used in this study has a dimension of $2.5\times2.5\times2.5$ cm$^3$ with all surfaces polished. The doping concentration of Na$^+$ is about 0.02%. The schematic of the experimental set-up for γ and α test is illustrated in Fig. 2. Two XP2020 PMTs were directly attached to the top and the bottom surfaces of the CsI(Na) crystal, while the other four sides are wrapped by the Enhanced Specular Reflector (ESR) film with a thickness of 65 μm. The PMTs signals are sent to oscilloscope and discriminators via NIM fan-out modules (model CAEN N625). The discriminator thresholds are set to be equivalent to 0.5 photoelectrons and the thermal noises of PMTs can then be effectively suppressed to 0.4 Hz by the coincidence of the two discriminator signals, which is set to have a width of 20 ns. The oscilloscope is TDS3054C with a sampling frequency of 5 GS/s and a memory depth of 2 μs (10000 points, 0.2 ns/point) for each of 4 channels. The data acquisition software is developed based on Labview8.6.

Using such a setup, the γ-ray test using a 2 micro-Curie $^{241}$Am source in front of the CsI(Na) crystal is performed. The source energy is a pure 59.5 keV line. The trigger is the coincidence of two PMTs attached to the CsI(Na) crystals and the trigger rate is about 10000 Hz while the event rate is about 0.7 Hz due to the low data transfer rate of the oscilloscope. The response of CsI(Na) to α is tested using the same setup while the source is replaced with 5.2 MeV α-ray from a 5 micro-Curie $^{239}$Pu source which is put close to the crystal surface. The crystal surface is grinded when testing the so-called surface effect of the crystal.

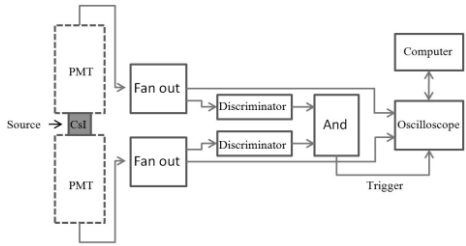

Fig. 2. Experimental set-up for the γ and α tests of CsI(Na) crystals.

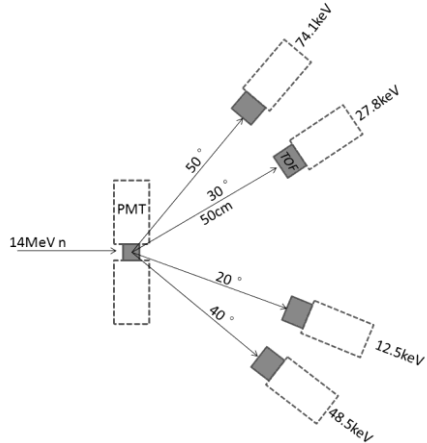

Fig. 3. Experimental set-up for the neutron beam test of CsI(Na) crystals.

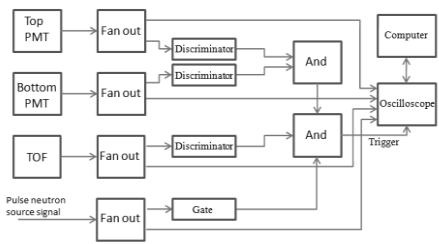

Fig. 4. Readout diagram of the CsI(Na) neutron beam test.

The response of the crystal to neutrons is studied by using a neutron gun at the China Institute of Atomic Energy (CIAE). It is a mono-energetic 14 MeV neutron beam generated from the D(T, α)n reaction, which is induced by a 250 keV deuteron beam impinging on a T-Ti target with a frequency of 1.5 MHz and pulse width of about 2 ns. There is a 1.5 m thick collimator wall, made of lead and concrete, between the source and the experimental set-up. In order to identify neutrons scattered from CsI(Na) against γ-ray backgrounds, a time of flight detector (TOF) was used as shown in Fig. 3. It is a Φ50x50 mm cylindrical plastic scintillator coupled with a XP2020 PMT and shielded with 1.5 cm thick lead. Fig. 4 shows the readout diagram of the neutron beam test. The trigger is the coincidence of the neutron source signal, the TOF signal and two PMT signals of the CsI(Na) crystal. In order to reduce the γ backgrounds induced by the neutron beam, the width of gate from the pulse neutron source signal is set to 100 ns and that from the logic unit of two PMTs attached to CsI(Na) crystals is set to 40 ns. Trigger timing is determined by the TOF detector and the trigger point is set at 200 ns for the oscilloscope. The waveforms of the four signals are directly acquired by the 4 channels oscilloscope simultaneously and then uploaded to the computer through Ethernet. The equivalent electron energy scale is calibrated by 59.5 keV γ-rays from [241]Am source, which is triggered by the coincidence of two PMTs attached to the CsI(Na) crystals.

**3. Neutron identification**

In order to identify neutron-induced events, we require a coincidence among the neutron source signal, the CsI(Na) crystal and the TOF detector. Furthermore, neutron scattering events can be selected by the flight time of 14 MeV neutrons scattered from the CsI(Na) crystal to TOF, which is about 10 ns while for γ it is only about 1.7 ns for the flying distance of 50 cm. Fig. 5 presents the TOF spectrum of neutrons and γs. The right peak is neutron while the left small peak is γ. The long tail of the neutron peak is due to slower neutrons from multiple scattering and γs from background. The cut for selecting neutron events is set to 12-25 ns. Backgrounds of γ will be further excluded by the energy spectrum as mentioned below.

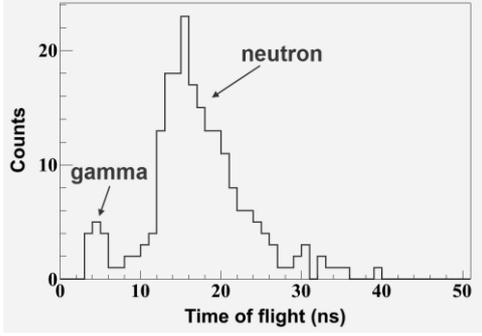

Fig. 5. TOF spectrum of neutron and γ. The right peak is neutron while the left small peak is γ.

The recoil energy of the nucleus ($E_r$) can be calculated by a simple kinematical equation using the incident neutron energy and the scattering angle of the neutron:

$$E_r = \frac{2A}{(1+A)^2}(1-\cos\theta)E_n \quad (1)$$

where $E_n$ is the neutron beam energy, $A$ is the masses of recoiling nucleus (Cs or I) and $\theta$ is the neutron scattering angle. The corresponding recoil energies of Cs are 12.5 keVr, 27.8 keVr, 48.5 keVr and 74.1 keVr at scattering angles of 20°, 30°, 40°, and 50°, respectively. The recoil energy of I (Iodine) is very close to these values. The measured energy should be proportional to the energy deposition if the energy linearity is good and could not exceed it. Furthermore, the spectrum of measured energy should have a peak in the expected energy range. These are the constraints from energy spectrum for neutron identification.

The measured energy spectrum of different nuclear recoils selected by the time of flight is presented in Fig. 6, which increases with the scattering angle as expected. In fact, it is not proportional to the energy deposition due to Quenching Effect, which will be discussed in Section 5. The long tail of ADC peak is mainly due to the γ backgrounds and the fraction is about 20% for ADC >300. Here the ADC count is the amplitude integration in a time window of 1.8 μs, namely from the trigger point to the end of the waveform. The pedestal of baseline is subtracted event by event, which is less than 3 ADC channels. A disproof is that there is no peak structure in the ADC spectrum of γs selected by PSD method described in Section 4.2.

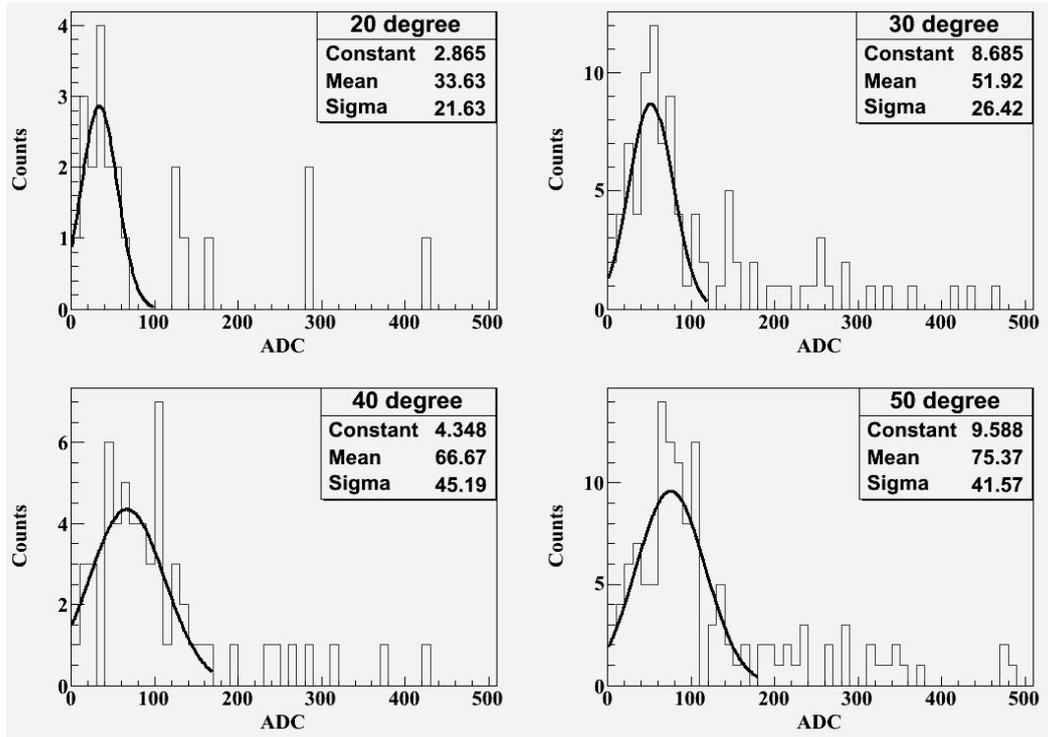

Fig. 6. Energy spectrums of nuclear recoils tagged by TOF. The corresponding recoil energy of Cs is 12.5 keVr, 27.8 keVr, 48.5 keVr, and 74.1 keVr, respectively.

## 4. Neutron-gamma discrimination

### 4.1 Waveform of γ/neutron

A typical waveform of the γ-ray (59.5 keVee) and the profile of 200 waveforms (50-60 keVee) from $^{241}$Am are shown in Fig. 7. The bin size of the waveform is 0.2 ns. Most of the signals have a rise time of about 50 ns and a fall time of about 600 ns.

In comparison, Fig. 8 presents a typical waveform of the identified neutrons (10 keVee) and the profile of 50 waveforms (5-10 keVee) at a scattering of angle $50^{\circ}$ as discussed in Section 3. Here, the energy scale is the equivalent electron energy calibrated by 59.5 keV γ-rays from $^{241}$Am source. The waveform difference is obvious: the decay time of γ signals is about 600 ns while that of nuclear recoil is only about 20 ns. The rise time of γ signals is about 50 ns while that of nuclear recoil is about 10 ns. This novel phenomenon is first discovered. Note that, there is a small fast peak with amplitude about 0.1 V at the start point ~ 200 ns of the γ-ray waveform and there is also a small part of slow components in the waveform of neutron. The scintillating mechanism will be discussed in Section 6.

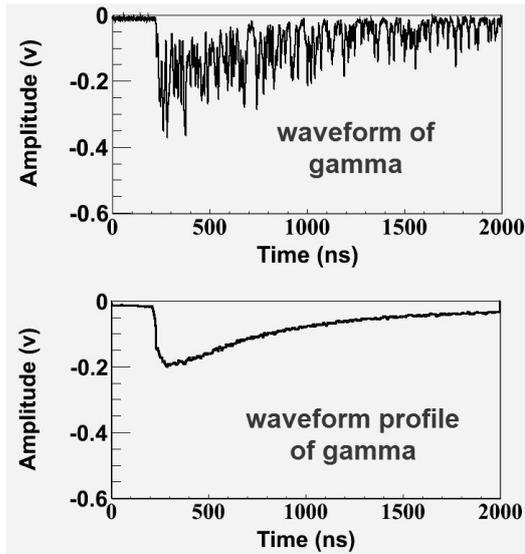

Fig. 7 A typical waveform of pure γ-ray (59.5 keVee) from $^{241}$Am and the profile of 200 waveforms (50-60 keVee).

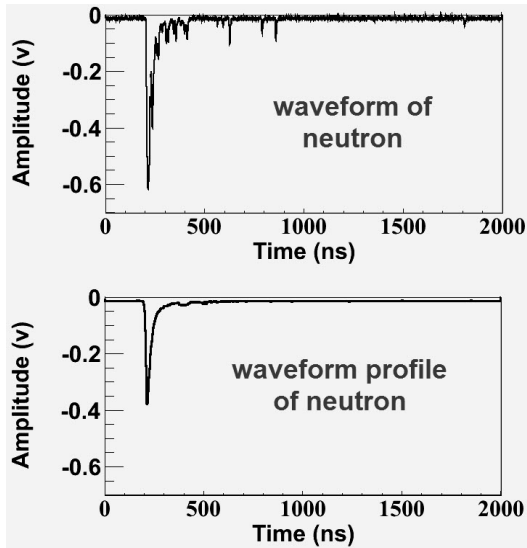

Fig. 8 A typical waveform of neutron (10 keVee) from neutron beam and the profile of 50 waveforms (5-10 keVee).

*4.2 PSD analysis*

Pulse Shape Discrimination (PSD) technique could be used for n/γ separation based on different waveforms of nuclear recoils and γ-rays discussed above, since the ratio of the fast to slow component is obviously different. Let's define ADC2 as the integration of the fast component, which is the amplitude integration in a window of 100 ns after the trigger point, and ADC as the amplitude integration from the trigger point to the end of the waveform. The scatter plot of the ratio, ADC2/ADC, for n and γ as a function of energy deposit is shown in Fig. 9. Neutrons and γs are well separated. Here we use the neutrons at a scattering angle of 50° and the γs from the calibration source $^{241}$Am.

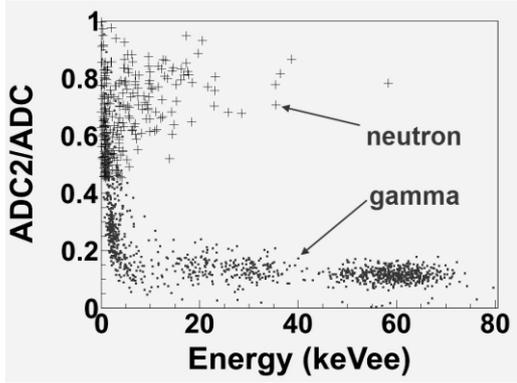

Fig. 9. Scatter plot of ADC2/ADC versus energy for n and γ. Dot is γ-ray from $^{241}$Am and plus is neutron at a scattering angle of 50°.

Let's define the rejection factor as the ratio of γ-ray events leaking into the nuclear recoil region to the total γ-ray events in the relevant energy range. This factor is actually the figure of merits for the n/γ separation and large statistics is required to determine it. Since the oscilloscope is too slow in data acquisition, another measurement was performed with a fast VME FADC system. The waveform sampling device is CAEN V1729 FADC with a 2 GS/s sampling frequency and 1.26 μs memory depth. A total of ~2 million γ-ray events from $^{241}$Am are acquired. Fig. 10 shows ADC2/ADC as a function of equivalent electron recoil energy. In the region of energy larger than 20 keVee and ADC2/ADC>0.65, there is only one event which is in the small round circle. It may be a mis-identified gamma, or an environmental fast neutron. Hence the rejection factor is $6.6 \times 10^{-7}$ or more, with ~80% efficiency for neutron nuclear recoil events, if we cut at ADC2/ADC>0.65 for events of energy >20 keVee. This result shows that CsI(Na) has the capability to distinguish γ backgrounds from nuclear recoil signals.

It is clear from Fig. 9 and 10 that the n/γ separation is not good at energies below 10 keVee. There are mainly two reasons: one is the statistical fluctuation due to too few photoelectrons; the other is the surface effect which will be discussed in Section 4.3. In fact, surface events can be excluded by vertex reconstruction hence the final rejection factor can be improved.

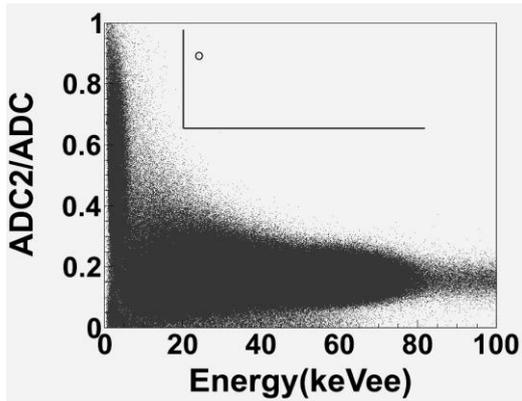

Fig. 10. Scatter plot of ADC2/ADC as a function of energy for γ-ray events. Only one event (in the small round circle) was seen in the region of ADC2/ADC > 0.65 and energy >20 keVee.

*4.3 Surface effect*

Surface effect is known for CsI(Na) crystal due to deliquescence, which will reduce Na$^+$ composition on the crystal surface [8]. We reveal that the so-called surface effect is one of the important unfavorable factors for n/γ separation in low energy region <10 keVee. We use 5.2 MeV α from $^{239}$Pu and 5.9 keV x-ray from $^{55}$Fe to test surface effect. The average projected range of different ions or particles injected into the CsI(Na) crystal are shown in Fig. 11 [9-11]. If the thickness of the surface layer $Na^+$ losing is more than 1 μm, $Na^+$ will be out of reach for α with energy less than 100 keV or electron with energy less than 10 keV. Hence the fluorescence of surface-α or electron is dominated by the luminescence of pure CsI. Pure CsI has a peak wavelength of 310nm and a decay time of 20 ns. Surface effect of x-ray with energy less than 10 keV has actually been observed [12]. We find that slow components appear obviously and fast components remain at the same time in the waveform of α when measured immediately after grinding off the old crystal surface, as shown in Fig. 12 and 13. The slow components will decrease when exposed to the air with time. We also find that the full energy peak of $^{55}$Fe (5.9 keV) can only be measured after grinding off the old crystal surface. These phenomena confirm the existence of surface effect. Surface effect will weaken the power of n/γ separation in low energy region <10 keVee, because the waveforms of surface x-rays are similar to neutron nuclear recoils. If the detector is a large crystal measured with 3D array of PMTs surrounding it, the surface events can be removed by vertex reconstruction. This can avoid the surface effect and achieve self-shielding from external background.

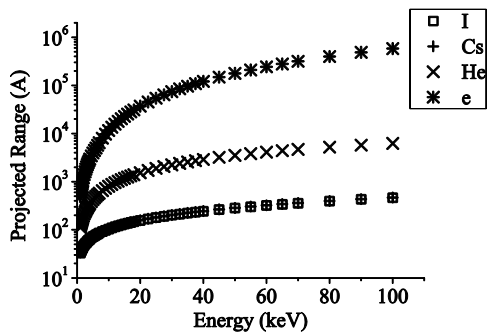

Fig. 11. The average projected ranges of different ions injected into CsI crystal.

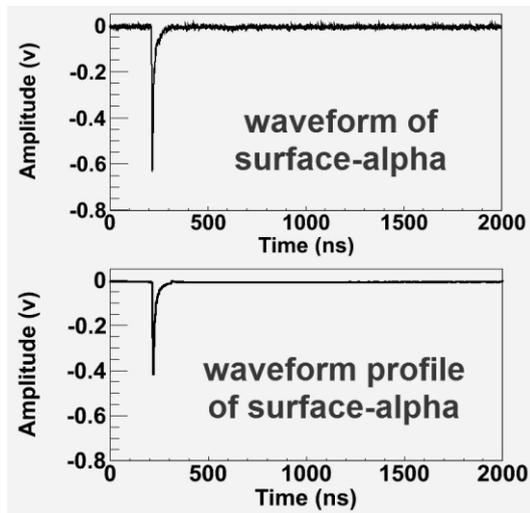

Fig. 12. A typical waveform of 5.2 MeV α from $^{239}$Pu and the profile of 200 waveforms with surface effect.

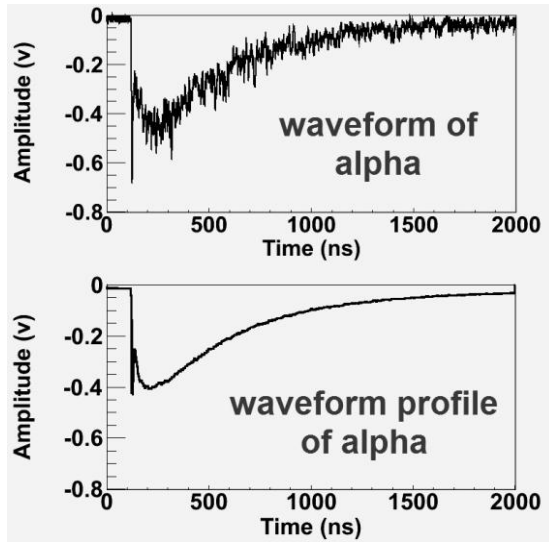

Fig. 13. A typical waveform of 5.2 MeV α from $^{239}$Pu and the profile of 200 waveforms without surface effect.

*4.4 Wavelength of fast and slow light*

It is discovered during the experiment that light of the fast and slow components have different wavelengths. This feature can be used to further distinguish neutron and γ, and help us to understand the scintillation mechanism for different incident particles. It is known that the peak wavelength of the slow component from CsI(Na) is 420 nm while that of the pure CsI is 310 nm. Hence, we use a filter made of the material UG11 with a 1 mm thickness. Its transmittance is more than 80% at 310 nm while less than 1% at 405 nm as shown in Fig. 14 [13]. One of the two PMTs, as shown in Fig. 2, is coupled to CsI(Na) through a filter. γ-rays from $^{241}$Am source and α-rays from $^{239}$Pu source are used to study the wavelength of the fast and slow components. Here we use surface-α to excite the fast light instead of neutron, since their fast light should have the same generation mechanism, which will be discussed in Section 6.

The ratio of amplitude integral with filter to that without filter (ADCfilter/ADC) can distinguish surface-α from γ-ray as shown in Fig. 15. It shows clearly that the fast light from surface-α has a peak wavelength of 310 nm, while the slow light from γ-ray has a peak wavelength of 420 nm. The wavelength of nuclear recoil should be the same as that of surface-α, hence resulting additional discriminating power between neutron and γ by using a wavelength filter.

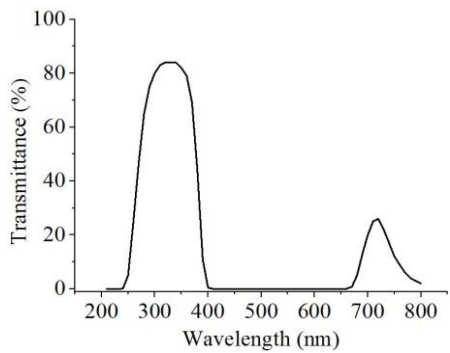

Fig. 14. Transmittance of UG11 as a function of wavelength.

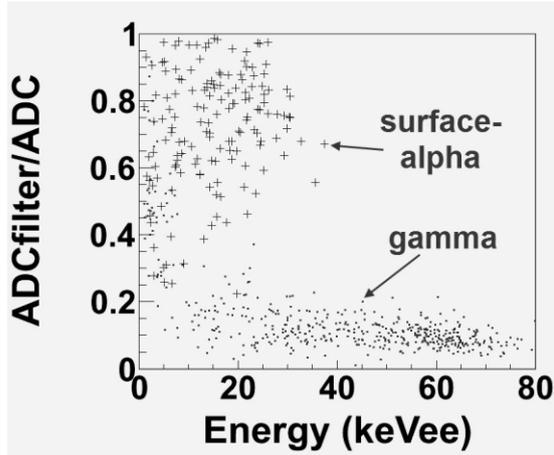

Fig. 15. Scatter plot of ADCfilter/ADC for γ and surface-α. Dot is γ-ray from [241]Am, and plus is surface-α from [239]Pu which is attenuated by a thin paper.

**5. Light yield**

The light yield is an important character of scintillator detectors. In particular, the PSD technique for n/γ separation needs sufficient photons to determine the pulse shape and reduce statistical fluctuations. Quenching Effect will reduce the light output for nuclear recoil.

Our measurement shows the light yield of γ-ray is 10-8 p.e./keVee at 1.8 μs shaping time in energy region 6-60 keVee, which is consistent with Ref. [14]. The recoil energy ($E_r$) is calculated with Eq. (1). The average measured energy ($E_{meas}$) for the TOF tagged events was determined by a Gaussian fit of the ADC spectrum as shown in Fig. 6 and calibrated by 59.5 keV γ-rays from [241]Am source. Using $E_{meas}$ and $E_r$, the quenching factor (QF) is calculated by

$$QF = \frac{E_{meas}}{E_r},$$

which is about 24-16% in the recoil energy region of 10-30 keVr.

**6. Discussion**

Different waveforms for γs and neutrons, as discussed above, demonstrate that CsI(Na) should have two different luminescence mechanisms. The slow components of 600 ns decay time may be associated with $Na^+$ [15] with a peak wavelength of 420 nm, while the fast components of 20 ns decay time may be from pure CsI with a peak wavelength of 310 nm. Such a luminescence mechanism can be explained by the combination of crystal structure, ionization density and thermal quenching effect. Average spacing of $Na^+$ is very large (~77 Å) in CsI(Na) crystal [8,16]. Ions with higher ionization density has a shorter projected range (covering less $Na^+$) than electron at the same energy as shown in Fig. 11. Hence the nuclear recoil has more fast components. It is known that pure CsI scintillates faster and emits less light than CsI(Na) because of serious thermal quenching. This suppresses the fast components of electron with lower ionization density. So the luminescence of neutron nuclear recoil is dominated by fast component and that of electron recoil is dominated by slow component.

**7. Conclusions**

Very different waveforms between nuclear recoils and electron recoils are found for CsI(Na) crystal. The luminescence mechanism could be related to different luminescence centers of $Na^+$ and pure CsI. The surface effect is compared and confirmed in our tests, and it must be avoided for better n/γ separation. By using the Pulse Shape Discrimination, the background rejection power can reach the order of $10^7$ for n/γ discrimination. CsI(Na) crystal is thus a good detector candidate for dark matter searches.

**Acknowledgments**

We are thankful to Sen Qian, Zhe Ning, Zaiwei Fu, Hongbang Liu and Yangheng Zheng for their support.